\documentclass[pdflatex,sn-mathphys-num]{sn-jnl}

\usepackage{comment}
\usepackage{graphicx}%
\usepackage{multirow}%
\usepackage{amsmath,amssymb,amsfonts}%
\usepackage{amsthm}%
\usepackage{mathrsfs}%
\usepackage[title]{appendix}%
\usepackage{xcolor}%
\usepackage{textcomp}%
\usepackage{manyfoot}%
\usepackage{booktabs}%
\usepackage{algorithm}%
\usepackage{algorithmicx}%
\usepackage{algpseudocode}%
\usepackage{listings}%
\usepackage{dsfont}

\theoremstyle{thmstyleone}%
\theoremstyle{thmstyletwo}%

\theoremstyle{thmstylethree}%

\begin{document}

\title[Article Title]{Quantum Frequency Resolved Optical Gating of Few-Cycle Squeezed Vacuum}


\author[1]{\fnm{Thomas} \sur{Zacharias}}

\author[2]{\fnm{Elina} \sur{Sendonaris}}

\author[1]{\fnm{Robert} \sur{Gray}}

\author[1]{\fnm{James} \sur{Williams}}

\author[1, 3]{\fnm{Ryoto} \sur{Sekine}}

\author[1]{\fnm{Maximilian} \sur{Shen}}

\author[1]{\fnm{Selina} \sur{Zhou}}

\author*[1, 2]{\fnm{Alireza} \sur{Marandi}}\email{marandi@caltech.edu}

\affil[1]{\orgdiv{Department of Electrical Engineering}, \orgname{California Institute of Technology}, \orgaddress{\city{Pasadena}, \postcode{91125}, \state{California}, \country{USA}}}

\affil[2]{\orgdiv{Department of Applied Physics}, \orgname{California Institute of Technology}, \orgaddress{\city{Pasadena}, \postcode{91125}, \state{California}, \country{USA}}}

\affil[3]{\orgname{PINC Technologies Inc}, \orgaddress{\city{Pasadena}, \postcode{91125}, \state{California}, \country{USA}}}




\abstract{
Offering terahertz of bandwidths and femtosecond timescales, ultrafast optics is enabling both the study of fundamental quantum optical phenomena and the advancement of quantum-enhanced applications.  
However, unlocking the full potential of ultrafast quantum optics requires accessing the temporal characteristics of ultrashort quantum pulses across ultrabroad bandwidths. 
This is particularly important in the near-infrared and visible range of the optical spectrum, which, unlike the terahertz and long-wave infrared, has remained beyond the reach of current techniques.
Here, we break this barrier by translating frequency-resolved optical gating (FROG), a widely used technique for ultrafast classical pulse characterization, to the quantum regime.
We show how such a quantum FROG can measure complex temporal modes and sub-optical-cycle quadrature covariances in the near-infrared, enabling complete characterization of microscopic Gaussian states. 
We experimentally use the quantum-FROG to report the measurement of quadrature correlations, complex temporal modes, and squeezing levels of multimode ultrafast squeezed vacuum states generated on a nanophotonic chip. 
We access multimode squeezing levels of a femtosecond quantum pulse approaching 7 dB and demonstrate FROG-based measurement bandwidths exceeding 100 THz.
Quantum FROG enables measurement of previously inaccessible quantum features of ultrashort pulses at the sub-optical-cycle regime and highlights a practical path to accessing terahertz of bandwidths in quantum optics for applications in computing, sensing, and imaging.  
}

\maketitle

Ultrafast optics offers substantial opportunities for quantum electromagnetic field measurements~\cite{riek2015direct, riek2017subcycle, benea2019electric}, spectroscopy and microscopy~\cite{herman2025squeezed, casacio2021quantum, siday2024all}, light-matter interactions~\cite{rasputnyi2024high, feist2015quantum, peller2021quantitative}, and information processing~\cite{jia2025continuous, roslund2014wavelength, kawasaki2025real} by providing access to terahertz-scale bandwidths. 
Such bandwidths have enabled the creation of entangled quantum states in the form of broadband frequency combs~\cite{jia2025continuous, roh2025generation, gwak2025completely, ra2020non, roslund2014wavelength, cai2017multimode, pinel2012generation, chen2014experimental, xie2015harnessing}.
Moreover, the large peak powers available in ultrashort pulses combined with dispersion-engineered nonlinear nanophotonic platforms provide an additional opportunity for enhanced nonlinear processes~\cite{nehra2022few}, placing non-Gaussian operations and states in the horizon~\cite{yanagimoto2022onset}. 

Despite substantial progress in a wide range of quantum optics experiments using ultrashort pulses \cite{paparelle2026experimental, ra2020non, cai2017multimode, roh2025generation, gwak2025completely, roman2024multimode, kalash2025efficient, serino2025programmable}, these experiments typically fall short of utilizing the temporal characteristics of  quantum fluctuations in the few-cycle and sub-cycle regime. Such a capability requires advancing the current measurement schemes which would greatly broaden the scope and scalability of ultrafast quantum optics. 
The most common techniques for characterization of ultrashort-pulse quantum states rely on projective measurements using pulse shaping, whose bandwidth and resolution are limited by the local oscillator and ultrafast pulse shaper~\cite{polycarpou2012adaptive, huo2020direct, roslund2014wavelength, medeiros2014full, paparelle2026experimental, ra2020non, cai2017multimode, roh2025generation, gwak2025completely, roman2024multimode}. 
Another scheme based on quantum pulse gating techniques additionally requires engineered optical nonlinearities~\cite{serino2025self, serino2025programmable}. 
In parallel, electro-optic sampling~\cite{benea2025electro} is also used for this purpose, but has been so far restricted to mid-infrared~\cite{riek2015direct, riek2017subcycle} and terahertz quantum pulses~\cite{benea2019electric} while requiring probe pulses shorter than the wavelength of interest. 

In contrast, for classical characterization of ultrashort pulses, FROG~\cite{trebino2000frequency} has been one of the most widely used techniques~\cite{kane2002characterization}, standing out for its experimental simplicity and high performance. 
FROG involves measuring a spectrogram followed by 2D phase retrieval to find the complex field profile of the ultrashort pulse. 
This technique has inspired an entire class of spectrographic measurement tools that have become the standard for classical ultrafast pulse characterization. 
Examples include the measurement of ultrabroadband pulses~\cite{gu2002frequency}, single cycle pulses~\cite{baltuvska1998amplitude, wirth2011synthesized}, ultraweak pulses~\cite{zhang2003measurement, fittinghoff1996measurement}, attosecond pulses~\cite{mairesse2005frequency}, and more~\cite{trebino2000frequency}.  


Using FROG for multimode quantum pulse characterization has remained challenging because it requires: (1) analytically formulating the quantum pulse spectrograms; (2) developing suitable phase-retrieval algorithms; (3) meeting measurement sensitivity requirements; and (4) accessing quadrature statistics of associated states. 
Although the analytical form of a quantum spectrogram has been observed to be distinct from that of a classical spectrogram, finding appropriate retrieval algorithms and measuring microscopic states has remained an outstanding problem~\cite{eto2024quantum}.
Phase retrieval techniques using single photon coincidence measurements~\cite{maclean2019reconstructing, davis2020measuring} have enabled ultrafast photon-pair characterization, but the experimentally simpler spectrographic techniques have remained elusive.
Importantly, these techniques~\cite{eto2024quantum, maclean2019reconstructing, davis2020measuring} do not access quadrature information required for continuous-variable quantum optics. 

Here, we overcome these challenges and introduce a quantum variant of FROG enabling high-bandwidth, high-resolution measurement of ultrafast squeezed vacuum. 
Our quantum FROG uses loss-tolerant phase-sensitive amplification \cite{shaked2018lifting, takanashi2020all} followed by a FROG-based spectrogram measurement and retrieval that enables recovering the quadrature covariances and temporal mode structure of quantum pulses. 
We experimentally demonstrate our technique by characterizing the temporal modes and squeezing levels of ultrashort pulse squeezed vacuum generated on a lithium niobate nanophotonic chip. 

Our work opens new avenues for quantum pulse measurement and temporal mode quantum state engineering for continuous variable quantum optics, while providing a practical and accessible tool to harness ultrafast quantum pulses for enhanced sensing and nonlinear optics.
\begin{figure*}[t] 
	\begin{centering}
    		\includegraphics[width=1\linewidth,trim={0cm 4.5cm 0.1cm 0.5cm},clip]{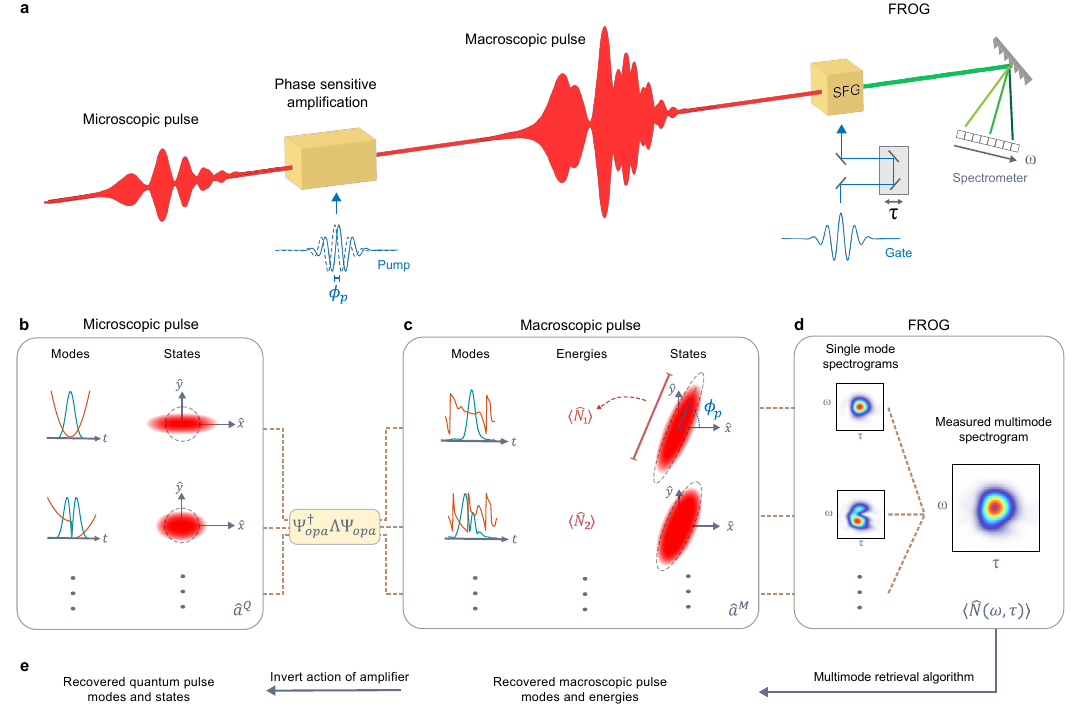}
    	\par\end{centering}
        \caption{ \textbf{Quantum FROG concept.} 
        \textbf{a} A microscopic squeezed vacuum pulse to be characterized is sent through a phase-sensitive amplifier, generating a macroscopic pulse whose spectrogram is measured using FROG.
        \textbf{b} The microscopic pulse is represented by its temporal modes and their Gaussian quadrature statistics, illustrated as phase space ellipses.
        \textbf{c} Phase-sensitive amplification transforms both the mode structure and quadrature statistics according to the amplifier mode transformation, $\Psi_{opa}$, gain transformation, $\Lambda$, and pump phase $\phi_p$. 
        The output macroscopic pulse has transformed temporal modes, mode energies (determined by the variance of the amplified quadrature), and Gaussian phase space distributions.     
        \textbf{d} Each mode has an associated single-mode spectrogram, and the sum of these spectrograms yields the measured multimode trace.  
        \textbf{e} A multimode phase-retrieval algorithm recovers the temporal modes and energies of the macroscopic pulse. 
        Applying the known amplifier transformation, $\Psi^{\dagger}_{opa}\Lambda\Psi_{opa}$ enables recovery of quantum pulse modes and quadrature statistics.  
    	}
	\label{fig: concept}
\end{figure*}
\section*{Concept}
Quantum FROG, displayed in Fig.~\ref{fig: concept}, includes two key physical processes: phase-sensitive amplification and frequency resolved optical gating.
The quantum pulse to be measured can be expressed as a collection of independent single-mode states living in distinct temporal modes.
Phase-sensitive amplification maps information about the temporal mode structure and quadrature statistics of the quantum pulse to the temporal mode structure and energy distribution of a macroscopic multimode pulse.
The generated macroscopic pulse can also be expressed as a collection of independent multimode states.
Due to the independent nature of the states, the FROG spectrogram generated by such a multimode pulse is equivalent to the sum of the associated single-mode spectrograms. 
This physical constraint is used to implement a phase-retrieval algorithm that enables the reconstruction of the temporal mode profiles and energies of the macroscopic pulse.
The mapping of information between the properties of the macroscopic pulse and quantum pulse is determined by the characteristic mode and gain transformations of the amplifier. 
Amplifying vacuum fluctuations followed by the FROG measurement and recovery procedure enables the reconstruction of these transformations, characterizing the phase-sensitive amplifier.  
Calibrating for the transformations enables quantum pulse reconstruction.
\section*{Theory}
We consider ultrafast squeezed vacuum, a multimode, zero-mean field, Gaussian quantum pulse whose structure is fully determined by its temporal modes and their quadrature covariances.
We describe the key relations connecting quantum pulse, macroscopic pulse, and the experimentally measured spectrogram that enable complete quantum pulse reconstruction. 

We represent the microscopic quantum pulse by its annihilation operators 
\begin{equation}
    \mathbf{\hat{a}}^\text{Q} = \mathbf{\hat{b}}^\text{Q}\Psi^\text{Q},
\end{equation}
where $\mathbf{\hat{a}}^\text{Q}$ and $ \mathbf{\hat{b}}^\text{Q}$ are row vectors containing annihilation operators in the temporal and principal (eigenmode) basis respectively. 
The rows of matrix $\Psi^\text{Q}$ contain the principal temporal mode functions, while the quantum statistics reside in the quadrature variances of $\mathbf{\hat{b}}^\text{Q}$. 

A phase-sensitive optical parametric amplifier (OPA)~\cite{wasilewski2006pulsed} maps the microscopic quadrature fluctuations onto measurable photon number in the macroscopic field.    
The OPA acts diagonally in its characteristic basis, characterized by unitary matrix $\Psi_{\text{opa}}$, in which each mode experiences independent quadrature-selective amplification. The amplified quadrature is determined by the pump-signal relative phase $\phi_p$. 
Expressing the annihilation operators of the quantum pulse in this basis as 
$\mathbf{\hat{c}}^\text{Q} = \mathbf{\hat{a}}^\text{Q} \Psi_{\text{opa}}^{\dagger}$, the amplified quadrature operator is $\mathbf{\hat{x}}^\text{Q} (\phi_p) = e^{-i\phi_p} \mathbf{\hat{c}^\text{Q}} + e^{i\phi_p} \mathbf{\hat{c} ^{\text{Q},\dagger}}$.
The macroscopic output field is then,
\begin{align}
     \mathbf{\hat{a}}^\text{M} &= \mathbf{\hat{x}}^\text{Q} (\phi_p) \Lambda {\Psi}_\text{opa},
    \label{eq:operatoropa}
\end{align} 
where $\Lambda$ is the diagonal gain matrix containing mode-dependent amplification values

We measure a spectrogram using sum-frequency-generation cross-FROG (SFG-XFROG), where the spectrum of the sum frequency signal between the macroscopic pulse and a classical gate pulse is recorded as a function of their relative delay. 
Because the amplified field remains Gaussian, it is separable into its principal basis, making it possible to express the measured SFG-XFROG  spectrogram, $\langle\hat{N}(\omega, \tau)\rangle$, as an incoherent sum of single-mode spectrograms,
\begin{equation}
    \langle\hat{N}(\omega, \tau) \rangle  \propto \sum_{m} \langle \hat{b}^{\dagger, \text{M}}_m \hat{b}^\text{M}_m \rangle | \mathcal{F} \{G(t-\tau )  \psi^M_m(t) \}|^2.
    \label{eq: ZMFieldSpectrogram}
\end{equation}
Here $\psi^M_m(t)$ and $\langle \hat{b}^{\dagger, \text{M}}_m \hat{b}^\text{M}_m \rangle$ denote the temporal profile and mean photon number of principal mode $m$ of the macroscopic pulse, and $G(t-\tau)$ represents the classical optical gating pulse.

In classical SFG-XFROG, an iterative two-dimensional phase retrieval algorithm is used to reconstruct a classical pulse envelope from a measured spectrogram.
While standard algorithms would work on a single-mode spectrogram, they fail on the multimode spectrogram considered here. 
By incorporating the constraints for a separable multimode state, we develop a custom retrieval algorithm (see Methods for details) enabling recovery of the complex temporal modes, $\psi_m(t)$, and mean photon numbers, $\langle \hat{b}^{\dagger}_m \hat{b}_m \rangle$.
These recovered quantities specify the macroscopic temporal correlation matrix, $\langle \mathbf{\hat{a}}^{\text{M,} \dagger} \mathbf{\hat{a}}^\text{M} \rangle$.


Using the amplification relation, this macroscopic correlation matrix relates to the quantum quadrature correlation matrix:   
\begin{align}
     \langle \mathbf{\hat{a}}^{\text{M,} \dagger} \mathbf{\hat{a}}^\text{M} \rangle &= \Psi^{\dagger}_\text{opa}  \Lambda \langle \mathbf{\hat{x}}^\text{Q} (\phi_p)^T \mathbf{\hat{x}}^\text{Q} (\phi_p) \rangle \Lambda {\Psi}_\text{opa}. 
    \label{eq:quadCorr1}
\end{align} 
To determine the unknown amplifier transformations, $\Lambda$ and $\Psi_\text{opa}$, we amplify vacuum ($\langle \mathbf{\hat{x}}^\text{Q} (\phi_p)^T \mathbf{\hat{x}}^\text{Q} (\phi_p) \rangle= \mathds 1$), measure the corresponding spectrogram, and retrieve its modes and energies. 
This calibration enables inverting for the amplifier transformation in Eq. \ref{eq:quadCorr1} giving access to the quadrature correlation matrix of the quantum pulse in the amplifier basis. 
Diagonalizing this matrix yields the quadrature variances of the quantum pulse in its principal basis and the corresponding change-of-basis transformation.
For real basis transformations, this enables recovering $\Psi^Q$ (see Supplementary Information (SI) for detailed theoretical analysis). 

\section*{Experiment}

\begin{figure*}[t] 
	\begin{centering}
    		\includegraphics[width=1\linewidth,trim={0 3.7cm 0 0cm},clip]{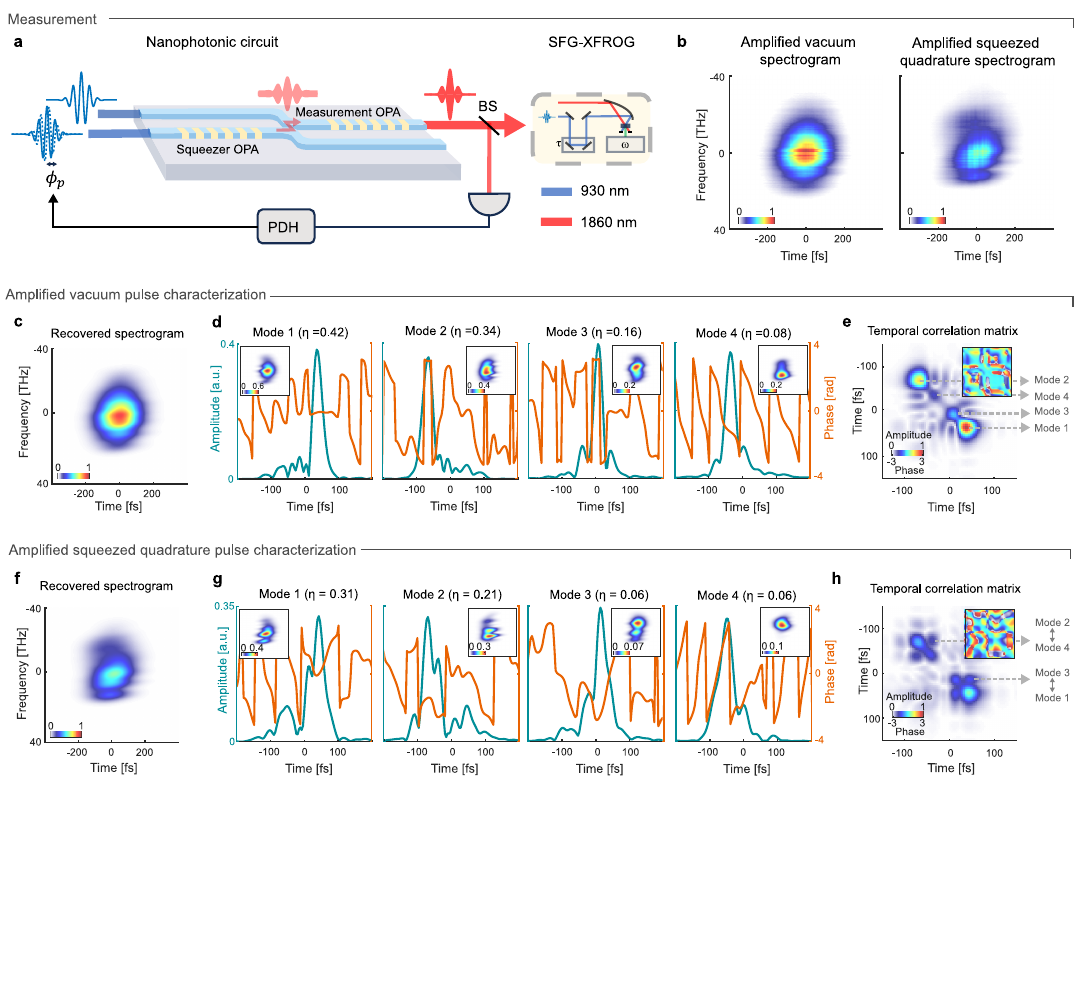}
    	\par\end{centering}
        \caption{ 
        \textbf{Measurement and characterization of macroscopic pulses.} 
        \textbf{a} Experimental scheme for ultrafast squeezed vacuum generation and phase-sensitive amplification on a lithium niobate nanophotonic chip.        
        The spectrogram of the amplified pulse is measured using an SFG-XFROG.  
        BS: High transmission beam splitter; PDH: Pound Drever Hall locking. 
        \textbf{b} Measured spectrograms of the amplified vacuum and amplified squeezed quadrature pulses, respectively. 
        \textbf{c, f} Recovered spectrograms for amplified vacuum and amplified squeezed quadrature pulses, respectively. 
        \textbf{d, g} Recovered complex orthonormal temporal modes. Each inset shows the single-mode spectrogram associated of that mode.  
        For both (d) and (g), $\eta$ is the mode energy normalized to the total amplified vacuum energy, providing a common reference across measurements. 
        \textbf{e, h} Temporal correlation matrices amplitude and phase (inset) for the amplified vacuum and amplified squeezed-quadrature pulses.  
        Double arrows indicate prominent intermode correlations. 
    	}
	\label{fig: macroscopicMeasurement}
\end{figure*}

\begin{figure*}[t] 
	\begin{centering}
    		\includegraphics[width=1\linewidth,trim={0 7.5cm 0 0},clip]{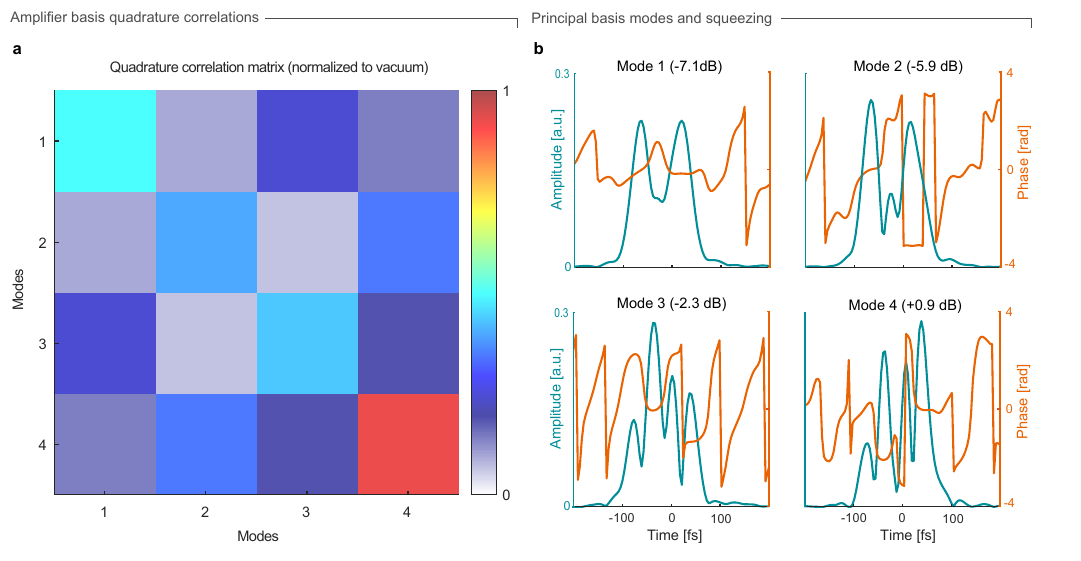}
    	\par\end{centering}
        \caption{ \textbf{Microscopic pulse characterization.} \textbf{a} Recovered squeezed quadrature correlation matrix of the quantum pulse in the amplifier basis.
        Diagonal terms give quadrature variances in this basis, and the off-diagonal terms represent inter-mode quadrature correlations. 
        \textbf{b} Recovered temporal modes and their squeezing levels in the principal basis. 
    	}
	\label{fig: quantumMeasurement}
\end{figure*}

Figure~\ref{fig: macroscopicMeasurement}a displays our experimental implementation of quantum FROG (see Methods for details). 
Our scheme utilizes two OPAs on a nanophotonic chip followed by an SFG-XFROG setup. 
The squeezer OPA generates a multimode squeezed state and the measurement OPA enables high-gain multimode phase-sensitive amplification. 
On-chip amplification enables tolerance to off-chip coupling losses~\cite{nehra2022few} that can limit the maximum measurable squeezing on the nanophotonic platform. 
We utilize a Pound-Drever-Hall locking scheme to control the relative phase between the pump and the multimoded squeezed state enabling quadrature selection.

We begin by measuring amplified vacuum and amplified squeezed quadrature spectrograms. 
We pump the measurement OPA with an ultrashort pulse at 930 nm and measure the spectrogram of the amplified vacuum fluctuations generated at 1860 nm. 
Figure \ref{fig: macroscopicMeasurement}b (Left) displays the corresponding spectrogram which indicates the shot-noise level and is used for characterization of the measurement OPA's modes and gain parameters. 
Next, the squeezer OPA is pumped to create a microsocopic squeezed vacuum state which is then amplified by the measurement OPA to a macroscopic state.   
The squeezed quadrature is selected by locking the pump phase to the trough of the detected signal, and the corresponding macroscopic pulse is measured using the SFG-XFROG.
Figure \ref{fig: macroscopicMeasurement}b (Right) displays the measured spectrogram of the amplified squeezed quadrature which has been normalized with respect to the shot-noise spectrogram.
A visual comparison of the two measured spectrograms confirms clear sub-shot noise intensity values characteristic of squeezing.

The amplified vacuum spectrogram is passed through the multimode phase retrieval algorithm and the recovered spectrogram is displayed in Fig. \ref{fig: macroscopicMeasurement}c. 
Figure \ref{fig: macroscopicMeasurement}d displays the recovered orthonormal complex temporal modes with their spectrograms inset and the fractional mode energy $\eta$.
Note that the sum of the inset spectrograms would yield the recovered spectrogram. 
Additionally, the complex orthonormal modes and the mode energies correspond directly to the amplifier mode and gain transformations, thus characterizing the measurement OPA. 
Using the temporal modes and energies, the complex temporal correlation matrix is calculated and displayed in Fig. \ref{fig: macroscopicMeasurement}e. 
The elements of this complex-valued matrix give information about the degree of correlation as well as the phase relationship between these two time instances.
The highly non-diagonal nature indicates correlations between different time instances. 
Note that most of the correlations exist within the temporal features corresponding to the individual modes and there are no prominent inter-mode correlations. 

Passing the amplified squeezed quadrature spectrogram through the retrieval algorithm yields the recovered spectrogram displayed in Fig. \ref{fig: macroscopicMeasurement}f, and mode information displayed in Fig. \ref{fig: macroscopicMeasurement}g. 
The displayed fractional mode energy, $\eta$, is still with respect to the total amplified vacuum energy for comparison purposes.  
The distinct orthonormal mode profiles and lower fractional mode energies of the amplified squeezed quadrature pulse with respect to amplified vacuum pulse carry signatures of the quadrature statistics and temporal mode structure of the microscopic squeezed vacuum pulse.
The amplified squeezed vacuum pulse temporal correlation matrix, displayed in Fig. \ref{fig: macroscopicMeasurement}h, reveals prominent inter-mode correlations missing in the amplified vacuum temporal correlation matrix.  
Additionally, the lower amplitude values for the amplified squeezed quadrature compared to the amplified vacuum temporal correlation matrix indicates time domain sub-shot noise characteristics.   

Using the squeezed quadrature temporal correlation matrix with the characterized measurement OPA gain and mode transformations enables quantum pulse reconstruction. 
Figure \ref{fig: quantumMeasurement}a displays the squeezed quadrature correlation matrix obtained by calibrating for the output amplifier temporal modes and gain transformations. 
This matrix carries information about quadrature correlations of the quantum pulse when projected to the basis of the measurement OPA. 
We note that all diagonal elements in this matrix (corresponding to quadrature variances in this basis) are below the vacuum level. 
The prominent non-diagonal terms reveal strong inter-mode quadrature correlations of the quantum pulse in this basis. 

Inverting the input temporal mode transformation of the amplifier reveals the temporal mode structure of the quantum pulse displayed in Fig. \ref{fig: quantumMeasurement}b. 
The squeezing level per mode was calculated from the diagonalized squeezed quadrature correlation matrix.
We measure 4 temporal modes of the quantum pulse with squeezing levels of $\{-7.1, -5.9, -2.3, +0.9\}$ dB. 
The first 3 modes are observed to have quadrature variances well below shot noise and the fourth mode is observed to have a quadrature variance greater than vacuum level.  

\begin{figure*}[t] 
	\begin{centering}
    		\includegraphics[width=1\linewidth,trim={0 11cm 0.5cm 0},clip]{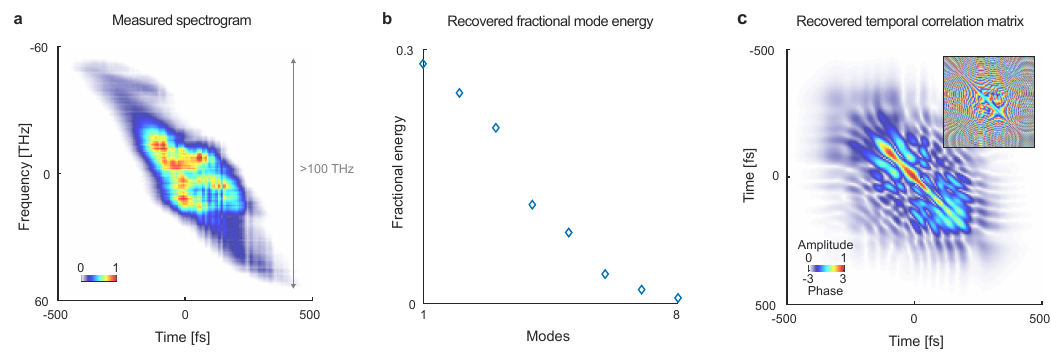}
    	\par\end{centering}
        \caption{ \textbf{Ultrabroadband measurement and recovery} \textbf{a} Measured spectrogram of ultrabroadband macroscopic pulse. \textbf{b} Recovered mode energy normalized to the total energy \textbf{c} Recovered temporal correlation matrix amplitude and phase (inset).
    	}
	\label{fig: ultrabroadband}
\end{figure*}

To highlight the high-bandwidth capabilities of FROG, we experimentally generate and measure an ultrabroadband pulse obtained by amplifying vacuum fluctuations (see Methods for experimental details). 
For ultrabroadband generation, we pump a lithium niobate nanophotonic OPA that had been dispersion-engineered to support near-single-cycle pulses \cite{gray2025two}.
The measured SFG-XFROG spectrogram is displayed in Fig. \ref{fig: ultrabroadband}a, showcasing spectral content spanning more than 100 THz. 
Multimode retrieval yields an 8-mode pulse, with the fractional energy per mode displayed in Fig. \ref{fig: ultrabroadband}b. 
Figure \ref{fig: ultrabroadband}c displays the recovered temporal correlation matrix revealing prominent non-diagonal temporal correlations.  

\section*{Discussion}
Our results close a long-standing gap between the tools routinely used for classical ultrafast pulse characterization and what has been available in the quantum regime.

In close analogy to the family of classical FROG variants, the presented technique can be extended to work with other spectrographic schemes.
Combining the retrieval technique with established classical FROG-based attosecond pulse characterization schemes~\cite{mairesse2005frequency} could enable reconstruction of quantum fluctuations at the attosecond timescale. 
Utilizing retrieval techniques based on ptychographic~\cite{witting2020retrieval} and learning based algorithms~\cite{zeng2024deep} may offer benefits in retrieval accuracy/time complexity. 
Moreover, nanophotonic FROG implementations~\cite{zacharias2025energy, yu2023frequency}, where the optical nonlinearity is on the nanophotonic platform, demonstrates that this approach is naturally suited to integrated platforms. 

The measurement and recovery techniques presented here can measure the temporal modes, mean photon numbers, and quadrature covariances of zero-mean field separable quantum pulses. 
Non-zero-mean field Gaussian states such as squeezed coherent states would have a different analytical formulation for the spectrogram requiring an adapted retrieval algorithm.
For biphoton characterization, the SFG-XFROG technique presented here can measure complex-valued joint spectral amplitude, Schmidt modes, Schmidt-coefficients, spectral purity, and Schmidt number (see SI). 
While the techniques presented here can measure the temporal modes and quadrature covariances of separable non-Gaussian states~\cite{ra2020non}, they do not have access to the higher order correlations required to characterize non-Gaussian states completely. 
Bringing single-shot FROG techniques~\cite{kane1993single, zhang2004sub}, well developed in the classical domain, suggests routes to measuring shot-to-shot fluctuating quantum fields enabling access to higher order correlations required for multimode non-Gaussian pulse measurements.   

The OPA-based mapping from microscopic quadrature fluctuations to the macroscopic state realizes a specific class of measurement-induced graph states (see SI) in the temporal domain \cite{cai2017multimode, roh2025generation}. 
Engineering the pump profile \cite{arzani2018versatile} and phase-matching function \cite{hurvitz2023frequency} opens a new path for large-scale temporal mode entanglement engineering. 
This has the advantage of broadband operation enabled by the nonlinear interaction, unavailable in current schemes \cite{cai2017multimode, roh2025generation, gwak2025completely, jia2025continuous}. 
Additionally, widespread availability of FROG can be exploited for pulsed sub-shot-noise sensing, such as nonlinear microscopy \cite{casacio2021quantum} and nonlinear interferometery \cite{ou2012enhancement}, as well as for applications leveraging the anti-squeezed quadrature for enhanced nonlinear conversion efficiencies \cite{rasputnyi2024high}.


\section*{Conclusion}
We have introduced and experimentally demonstrated FROG for characterizing ultrafast squeezed vacuum.
By combining phase-sensitive amplification on a nanophotonic OPA with an SFG-XFROG and a multimode retrieval algorithm, we retrieve complex principal modes, fractional mode energies, and temporal-basis correlation matrices of an amplified quantum pulse. 
Calibrating for the OPA allows us to reconstruct the squeezed quadrature correlation matrix and to recover four temporal modes with squeezing levels -7.1, -5.9, -2.3, +0.9 dB on chip.
Given its experimental simplicity, compatibility with integrated platforms, and intrinsic high bandwidth and resolution, quantum FROG provides a versatile measurement tool and naturally interfaces with applications in temporal-mode quantum information processing, nonclassical sensing, and nonlinear quantum optics.



\bibliography{references}

\newpage
\section*{Methods}

\subsection*{Multimode retrieval algorithm}
The classical SFG-XFROG uses a 2D phase retrieval algorithm to recover the complex pulse profile, $P(t)$, from a spectrogram, $I(\omega, \tau) = |\mathcal{F}\{G(t-\tau) P(t) \}|^2$.
Here $\tau$ indicates the delay between gate pulse $G(t)$ and unknown pulse $P(t)$ and $\mathcal{F}$ Fourier transforms from time $t$ to frequency $\omega$. 
The spectrogram of a separable pulse (Eq. \ref{eq: ZMFieldSpectrogram}) is mathematically distinct and therefore cannot be used with standard 2D phase-retrieval algorithms, necessitating a custom algorithm. 
We observe that the spectrogram defined by Eq. \ref{eq: ZMFieldSpectrogram} is mathematically analogous to the equation defining the spectrogram generated by a classical train of non-repeating pulses \cite{escoto2019retrieving}. 
This allows us to utilize phase retrieval techniques that have recently been developed for noisy pulse measurements  \cite{rhodes2013pulse, escoto2019retrieving, haham2017multiplexed, witting2020retrieval, bourassin2015partially}. 
We develop a custom recovery algorithm, Separable State Generalized Projections Algorithm (SSGPA), that builds on the Mixed-State Generalized Projections Algorithm (MSGPA) \cite{bourassin2015partially}. 
MSGPA enable phase retrieval of a spectrogram, $S_{meas}(\omega, \tau)$, that is generated as an incoherent sum of spectrograms, $S_{meas}(\omega, \tau) = \sum_k^N|s_k(\omega, \tau)|^2$. In each iteration of MSGPA, the algorithm reconstructs $N$ spectrograms in parallel and applies a data constraint, $\sqrt{S_{meas}/\sum_k|s_k|^2}$ per spectrogram that ensures the spectrogram equation validity. 
SSGPA incorporates an orthogonality constraint enforced by applying the Gram-Schmidt process in each iteration to the complex pulses recovered by the MSGPA.
Additionally, SSGPA works with normalized modes, keeping track of the normalization constant. 
Thus SSGPA is specifically designed to work with Eq. \ref{eq: ZMFieldSpectrogram}, enabling the recovery of mean photon number per mode and the complex temporal mode profile.
We note that we use mean photon number per mode and mode energy interchangeably in this work, as they are proportional for a fixed optical carrier frequency. 

Additional details can be found in SI. 

\subsection*{Experiment}
Design and fabrication details about the device used for the results presented in Fig. \ref{fig: macroscopicMeasurement} and \ref{fig: quantumMeasurement} can be found in \cite{nehra2022few, williams2025ultrafast}.
Here, we elaborate on the measurement apparatus displayed in Fig. \ref{fig: macroscopicMeasurement} a. 
The optical pumps and gate used in the experiment are derived from a 100-fs TiS mode-locked laser operated at a central wavelength of 930 nm. 
The laser is split into three paths, with two paths going to the chip and the third path going to the SFG-XFROG.  
The two paths to the chip are aligned to the squeezer and measurement OPA waveguides. 
The squeezer optical path contains a moving mirror that sits on a piezo-electric actuator for tuning the pump phase. 
Light is coupled on and off the chip using reflective objectives to prevent distortions of the pump and signal temporal profiles.
Pumping the squeezer OPA generates a microscopic multimode squeezed state at 1860 nm, which couples to the measurement amplifier through an adiabatic coupler. 
Temporally overlapping the measurement pump with the squeezed state results in phase-sensitive-amplification, generating a macroscopic pulse at 1860 nm. 
The macroscopic pulse is passed through a long-pass filter that filters out the pump. 
The resulting signal has $\langle \hat{N}\rangle \approx 10^6$, ten percent of which is directed to a detector using a high transmission beam splitter. 
The detected signal is fed back to the piezo-electric actuator through a PDH locking scheme and locked to the signal trough, corresponding to the squeezed quadrature. 
The transmitted light enters the SFG-XFROG and is gated on a 1 mm Lithium Iodate nonlinear crystal. 
The gate pulse energy is 5.2 nJ. 
Spectrum of the upconverted light is measured using a compact CMOS spectrometer (Thorlabs CCS200) for different gate delays controlled by a motorized linear translation stage.  
The spectrometer integration time was 200ms and 300ms for the amplified vacuum spectrogram and amplified squeezed vacuum spectrogram measurements respectively. 
The measured spectrograms are downsampled to a 512x512 grid and scaled to the same intensity axis (by accounting for the difference in integration time) before being processed by the multimode phase retrieval algorithm. 
The RMS loss between the measured and recovered spectrograms normalized to their peak intensities are $0.003$ and $0.004$ for amplified vacuum and amplified squeezed quadratures respectively.

Design and fabrication details about the device used for the results presented in Fig. \ref{fig: ultrabroadband} can be found in \cite{gray2025two}.
An ultrafast macroscopic pulse is generated by amplifying vacuum fluctuations using a broadband lithium niobate nanophotonic parametric amplifier. 
A macroscopic pulse, centered at 2090 nm, is generated by pumping the device with 100-fs pulses centered at 1045 nm. 
The spectrogram is then measured using the SFG-XFROG using a 100-fs gate pulse at 1045 nm. 
We use a thin $100\mu m$ Lithium Iodate crystal that enables broadband FROG operation.
The measured spectrogram is downsampled to a 256x256 grid before being processed by the multimode phase retrieval algorithm. 
The RMS loss between the measured and recovered spectrograms normalized to their peak intensities is $0.02$. 

Additional details can be found in SI. 

\backmatter

\bmhead{Acknowledgements}
The authors thank Nicolas Englebert and Rithvik Ramesh for fruitful discussions. 
The device nanofabrication was performed at the Kavli Nanoscience Institute (KNI) at Caltech. The authors gratefully acknowledge support from DARPA award D23AP00158, ARO grant no. W911NF-23-1-0048, NSF grant no. 2408297, 1918549, AFOSR award FA9550-23-1-0755, the Center for Sensing to Intelligence at Caltech, the Alfred P. Sloan Foundation, and NASA/JPL.

\bmhead{Author contributions}
T.Z., E.S., R.G, and A.M. conceived the idea. T.Z. conducted the theoretical analysis, developed the retrieval algorithm, and performed the experiments. 
E.S. assisted with the theoretical analysis. 
R.G. assisted with experiments and theoretical analysis. 
J.W. assisted with the experiments. 
R.S. fabricated the chip illustrated in Fig.\ref{fig: macroscopicMeasurement}.
M.S., S.Z., and J.W. fabricated the chip used for the result displayed in Fig. \ref{fig: ultrabroadband}.
T.Z. wrote the manuscript with input from all authors. 
A.M. supervised the project. 

\bmhead{Data availability}
The data that support the findings of this study are available from the corresponding author upon reasonable request. 

\bmhead{Code availability}
Custom code used for data analysis and mode retrieval is available from the corresponding author upon reasonable request. 

\bmhead{Supplementary information}
The supplementary information includes details about theory, recovery algorithm, and analysis. 
It includes additional experimental measurements FROG measurements. 

\bmhead{Competing financial interests}
T.Z., E.S., R.G., and A.M., are inventors on a provisional patent application CIT-9310-P.  
R.S. and A.M. are involved in developing photonic integrated nonlinear circuits at PINC Technologies Inc.
R.S. and A.M. have an equity interest in PINC Technologies Inc.

\end{document}